\documentclass[journal=jacsat,manuscript=article]{achemso}
\setkeys{acs}{super = false}
\usepackage[version=3]{mhchem} 

\usepackage{xcolor}
\usepackage{soulutf8}
\usepackage{gensymb}
\usepackage{tablefootnote}
\usepackage[normalem]{ulem}
\usepackage[bb=stix]{mathalpha}
\usepackage{booktabs}
\usepackage{siunitx}
\usepackage{caption}
\usepackage{geometry}
\usepackage{array}
\usepackage{tabularx}
\usepackage{makecell}
\usepackage{mathrsfs}
\usepackage[allow-number-unit-breaks]{siunitx}
\setcitestyle{numbers,square, sort & compress}

\AtBeginDocument{%
}

\title{Phase-Field Simulation of Dendrite Evolution in All-Solid-State Sodium Batteries during Cycling}
\author{Chengyin Wu}
\affiliation[MSE]{Department of Materials Science and Engineering, The Ohio State University, Columbus, OH 43210}
\author{Wolfgang Windl}
\affiliation[MSE]{Department of Materials Science and Engineering, The Ohio State University, Columbus, OH 43210}
\author{Jung-Hyun Kim}
\affiliation[MAE]{Department of Mechanical and Aerospace Engineering, The Ohio State University, Columbus, OH 43210}

\author{Yanzhou Ji}
\email{ji.730@osu.edu}
\affiliation[MSE]{Department of Materials Science and Engineering, The Ohio State University, Columbus, OH 43210}

\begin{document}

\maketitle

\begin{abstract}
Dendrite growth during cycling remains a critical challenge for all-solid-state batteries (SSBs), limiting the full realization of their inherent safety and high energy density. In particular, the mechanisms of continuous dendrite penetration during charge-discharge cycling remain poorly understood and are difficult to characterize experimentally. This study applies a phase-field model, informed by density functional theory calculations, to rationalize and visualize the dendrite penetration behaviors during cycling in sodium (Na) SSBs with pure Na or Na-Sb alloy anodes and polycrystalline Na$_3$SbS$_4$ electrolyte. We show that dendrite stripping is intrinsically asymmetric with respect to plating due to grain boundary geometry, leading to the formation of isolated Na metal that persists between cycles. This residual Na metal becomes kinetically stabilized at grain-boundary junctions and is readily reactivated during subsequent plating, thereby accelerating and amplifying dendrite penetration. We further investigate the effects of applied voltage, solid-electrolyte microstructure, and anode chemistry on this phenomenon. These findings establish isolated Na metal as a key contributor for continued dendrite propagation in Na SSBs and provide design principles for stabilizing anode/electrolyte interfaces in Na SSBs.

\end{abstract}


\section{1. Introduction}
The increasing attention to all-solid-state batteries (SSBs) is closely associated with the rapid growth of electric vehicles (EVs), which require safer energy-storage technologies with higher energy density. In particular, lithium solid-state batteries (Li SSBs) have advanced significantly in recent years.\cite{janek2023challenges, zheng2024thermal} Although Li SSBs offer promising advantages including high energy density and excellent ionic conductivity, their large-scale commercialization remains constrained by high production costs, environmental concerns associated with lithium mining, and long-term sustainability challenges.\cite{vera2023environmental} In contrast, sodium solid-state batteries (Na SSBs) offer a more cost-effective and sustainable alternative, because of the global abundance of sodium.\cite{zhao2018solid} When paired with Na-ion-conducting sulfide solid electrolytes, Na SSBs can achieve remarkable ionic conductivities up to 32~$\text{mS/cm}$ at room temperature.\cite{hayashi2019sodium} Despite these advantages, Na SSBs still face critical issues similar to Li SSBs, most notably dendrite growth during electrochemical cycling.\cite{lee2019sodium} Therefore, advancing our understanding of the transport behavior , stability, and failure mechanisms Na SSBs is essential for their development as sustainable next-generation energy-storage systems.

A comprehensive understanding of failure mechanisms in SSBs requires resolving the coupled mass transport, charge transfer, and interfacial dynamics at solid electrolyte (SE) interfaces. However, the inherently complex and rapid redox processes in Na SSBs make in situ characterization of microstructural evolution during charge-discharge cycling particularly challenging. Techniques such as high-resolution transmission electron microscopy (HRTEM)\cite{wang2012tracking, wei2025intermediate, liu2024situ}, electron backscatter diffraction (EBSD)\cite{fuchs2024imaging}, and nuclear magnetic resonance (NMR)\cite{zhang2024solid} have been employed to investigate microstructural changes. However, these methods remain limited by spatial or temporal resolution and experimental constraints, making it difficult to fully resolve nanoscale-to-microscale transport kinetics and interfacial processes. Computational modeling therefore provides a valuable complementary approach\cite{xu2021guiding,ling2022review,zhang2019progress}, for investigating nanometer-scale ion transport, deposition mechanisms, and the influence of external factors such as stackpressure and applied overpotential.

A wide range of computational strategies have been applied to solid-state batteries, including density functional theory (DFT)\cite{tian2019interfacial}, molecular dynamics (MD)\cite{ushirogata2015near, ren2024visualizing}, and kinetic Monte Carlo (KMC)\cite{takenaka2014electrolyte} simulations. Compared with these approaches, phase-field methods\cite{tian2019interfacial,liu2021dendrite,zhang2025simulating,wang2020application} are particularly well suited for describing mesoscale microstructural evolution over length and time scales relevant to electrochemical degradation. They can capture the coupled evolution of morphology, transport kinetics, and structure–property relationships without explicitly tracking moving interfaces. Phase-field models have been successfully employed to study dendrite growth\cite{mu2019numerical, tian2019interfacial, gao2024phase, gao2024controlling, pant2024dendrite, zhang2025isolated, chadwick2025suppression, chen2015modulation, tantratian2021unraveling}, solid electrolyte interphases (SEI) formation \cite{zhang2025simulating}, interfacial engineering strategies\cite{yurkiv2018phase}, and void formation\cite{barai2024study} in batteries. These studies demonstrate the potential of phase-field modeling provide mechanistic insights and guide solid-state battery design. 

Although phase-field models have been used to simulate charge-discharge cycling in liquid electrolyte batteries\cite{pant2024dendrite,han2025combination}, they are rarely applied to SSBs. In this work, we employ DFT-informed phase-field simulations to investigate dendrite growth during repeated plating–stripping cycles in solid-state batteries. As a representative system, we consider Na metal anodes paired with the polycrystalline sulfide solid electrolyte \ce{Na3SbS4} (NAS). Simulations are performed under varying applied voltages and grain structures to capture the interplay between electrochemical and geometric factors. Through this study, we aim to reveal the mechanisms governing microstructural evolution during repeated plating and stripping and to provide guidance for optimizing Na-SSB design toward stable and reliable device performance.

\section{2. Methods}
In this study, we rely on phase-field simulations to predict Na dendrite evolution, while using DFT calculations to get materials-specific parameters for the phase-field model. In particular, we predict surface electron concentrations in NAS from DFT. As noted in \cite{tian2018computational}, surface electrons play a critical role in dendrite formation and evolution in solid-state batteries, where excess electrons at surfaces and grain boundaries (GBs) of polycrystalline solid electrolytes can promote additional electrodeposition. This phenomenon reduces the resistance of solid electrolytes to dendrite growth, thereby compromising electrochemical stability.\cite{tian2019interfacial} We note that the SE structures considered here do not contain explicit free surfaces but are dominated by grain boundaries and internal interfaces. Therefore, the use of surface electron densities derived from DFT slab calculations to represent interfacial electron densities constitutes an approximation.

\subsection{2.1 DFT Calculations}

The DFT calculations were performed using the Vienna Ab initio Simulation Package (VASP) \cite{kresse1996efficiency, kresse1996efficient}. The NAS crystal structure (mp-10167) was sourced from the Materials Project database.\cite{horton2025accelerated,jain2013commentary} To investigate excess surface electrons, slab models were constructed for the (100), (110), and (111) surfaces, where all cells have a vacuum of at least 35~{\AA}. Exchange-correlation effects were treated within the generalized gradient approximation (GGA) using the Perdew, Burke, and Ernzerhof (PBE) functional.\cite{perdew1996generalized} A plane-wave cut-off energy of 300 eV was used. For the bulk and slab systems, Brillouin zone integration was performed using $\Gamma$-centered k-point grids of $3 \times 3 \times 3$ and $3 \times 3 \times 1$, respectively. To determine the excess electron density at the temperature of $T$ = 300~K used in the phase-field simulations, Fermi-Dirac smearing was applied to account for the electronic occupation while maintaining a fixed ionic configuration.

\subsection{2.2 Phase-field Model}
Consider an electrochemical reaction of ${\rm Na}^{+} + e^{-} \longleftrightarrow {\rm Na}$ in the Na-SSB system without any side reactions, mechanical effects, and compounds forming during plating and stripping. To describe the microstructure evolution during this process, we define an order parameter $\xi$ to distinguish Na metal ($\xi=1$) and NAS SE ($\xi=0$), and another order parameter $\phi_{g}$ to describe SE grains, distinguishing SE grain interior ($\phi_g=1$) and GBs ($\phi_g=0$)\cite{tian2019interfacial}. We also define $c_{\rm Na}$ as the total site concentration of Na, $c_0$ as the reference Na site concentration of NAS SE, $c_{\mathrm{Na}^+}$ as the local Na ion concentration, and $c_{e^-}$ as the electron concentration. $\tilde{c}$ =$\frac{c}{c_{\rm Na}}$ corresponds to the order parameter $\xi$, and $\tilde{c_{i}} = \frac{c_i}{c_0}$ ($i={\rm Na}^+, e^-$) stands for the reduced Na ion and electron concentrations. The total free energy of this system can be written in the form of:

\begin{equation}
    \mathscr{F} = \int (f_{\rm bulk} + f_{\rm grad} + f_{\rm elect})dV 
\end{equation}

The bulk free energy is defined as:

\begin{equation}
    f_{\rm bulk} = wg^{\rm dw}(\xi) + 5w\sum_{\phi_g = 1}^N \xi^2\phi_g^2 +f_0(\tilde{c},\tilde{c}_{\mathrm{Na}^+},\tilde{c}_{e^{-}})
\end{equation}

where $w$ is the double-well potential height, $g^{\rm dw}$ is the double-well potential energy function defined as $g^{\rm dw}(\xi) = \xi^2(1-\xi)^2$; we assume $\phi_{g}$ is not evolving during Na plating and stripping.

$f_0$ is the chemical free energy of the system: \cite{chen2015modulation}

\begin{equation}
    f_0 = \sum_i c_i\mu_i^0 + c_{\rm Na}RT(\tilde{c}ln\tilde{c})+c_0RT(\tilde{c}_{\mathrm{Na}^+}ln\tilde{c}_{\mathrm{Na}^+} + \tilde{c}_{e^{-}}ln\tilde{c}_{e^{-}})
\end{equation}
where $\mu_i^0$ ($i={\rm Na}^+, e^-$, Na) is the reference chemical potential of species $i$. R is the ideal gas constant.

The gradient energy can be defined as:

\begin{equation}
    f_{grad} = \frac{1}{2}\kappa(\theta)(\nabla\xi)^2
\end{equation}

where $\kappa(\theta)$ is the orientation-dependent gradient coefficient with expression of $\kappa(\theta)=\kappa_{0}(1+\delta \times cos(\omega\theta(\xi_{y},\,\xi_{x})))$. $\kappa_0$ and $w$ are related to the Na/SE interfacial energy and interface thickness, $\delta$ is associated with the anisotropy strength of the Na/SE interface, $\omega=4$ is the crystallographic symmetry of body-centered cubic Na metal, and $\theta(\xi_{y},\,\xi_{x})$ denotes the local interface orientation angle defined by the direction of normal vector from the order-parameter gradients $(\xi_y,\xi_x)$, measured with respect to the reference $\xi_x$ axis.



The electrostatic energy density is:

\begin{equation}
   f_{\rm elect} = F\sum{z_ic_i}\phi
\end{equation}

where $F$ is the Faraday constant, $z_i$ ($i={\rm Na}^+, e^-$) is the valence of the species, and $\phi$ is the electric potential.

The total overpotential can be defined as:

\begin{equation}
    \eta = \Delta\phi- \Delta\phi^{\rm eq} = \frac{\Delta\mu}{nF} = \frac{1}{nF}\sum_i\frac{\partial\mathscr{F}}{\partial c_i}
\end{equation}

where $\Delta\phi^{\rm eq}$ is the reference overpotential and $n$ = 1 for Na charge transfer.





Therefore, the governing equation for electrodeposition in a solid-state half-cell is a nonlinear Allen-Cahn equation coupled with Butler-Volmer kinetics \cite{chen2015modulation, liang2012nonlinear}:

\begin{equation}
  \begin{split}
    \frac{\partial\xi}{\partial t} = &-L_{\xi}[wg'(\xi)+10w\xi\phi_g^2 -\nabla\cdot(\frac{\partial f_{\rm grad}}{\partial \nabla\xi}) ] \\
        &\quad -L_{\phi}h'(\xi)\exp\left[\left(\frac{(1-\alpha)F\eta}{RT}\right)-\tilde{c}_{\mathrm{Na}^+}\tilde{c}_{\mathrm{e}^-}\left(\frac{(-\alpha)F\eta}{RT}\right)\right]
  \end{split}
\end{equation}

where $L_\xi$ is the interface mobility, $L_\phi$ is associated with the exchange current of electrodeposition, $\alpha$ is the symmetry coefficient of the anodic and cathodic reactions, and $h(\xi)=6\xi^5-15\xi^4+10\xi^3$ is an interpolation function. $\tilde{c}_{e^{-}}$ has an expression of $\tilde{c}_{e^{-}} = \frac{c_{e^{-}}}{c_0} = \xi+\frac{c_{\rm surf}}{c_0}(1-\xi)(1-\phi_{g})$ where $c_{\rm surf}$ is the surface electron concentration of SE obtained from DFT calculations. The details of the nonlinear phase-field model for electrodeposition can be found in Chen et. al.\cite{chen2015modulation}

The governing reaction-diffusion equation for Na ion concentration evolution is:

\begin{equation}
    \frac{\partial\tilde{c}_{\mathrm{Na}^+}}{\partial t} = \nabla \cdot D^{\rm eff}(\nabla \tilde{c}_{\mathrm{Na}^+} + \frac{F\tilde{c}_{\mathrm{Na}^+}}{RT}\nabla\phi) - \frac{c_{\rm Na}}{c_0}\frac{\partial\xi}{\partial t}
\end{equation}

where $D^{\rm eff}$ is the effective diffusivity of ${\rm Na}^+$ in the form of $h(\xi)D_{\rm Na} + (1-h(\xi))\phi_{g}D_{\rm SE}$. $D_{\rm Na}$ and $D_{\rm SE}$ are the diffusivity in pure sodium and solid electrolyte, respectively. 

The electric potential distribution is governed by the current conservation equation:

\begin{equation}
    \nabla \cdot [\sigma^{\rm eff}(\xi)\nabla\phi] = -c_{\rm Na}F\frac{\partial\xi}{\partial t}
\end{equation}
where $\sigma^{e\rm ff}$ is the effective conductivity described as $\sigma^{\rm eff}(\xi) = h(\xi)\sigma_{\rm Na} + (1-h(\xi))\phi_{g}\sigma_{\rm SE}$ and $\sigma_{\rm Na}$ and $\sigma_{\rm SE}$ are the overall conductivity of pure sodium and solid electrolyte, respectively.

\begin{figure}[!h]
    \centering
    \includegraphics[width=\linewidth]{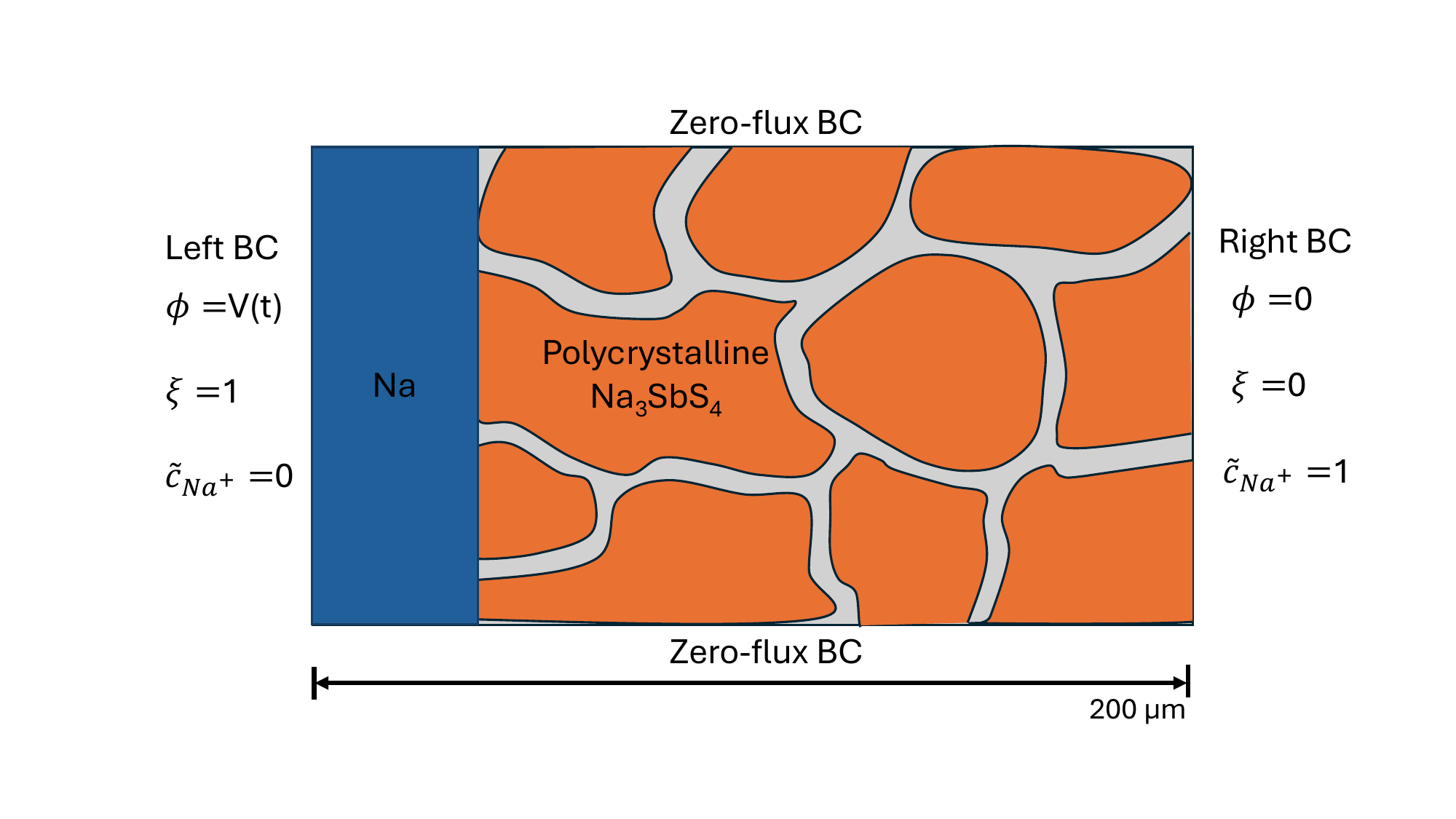}
    \caption{Schematic illustration of phase-field model setup and boundary conditions (BCs). The blue, orange and grey colors correspond to the Na anode, NAS grains, and NAS GBs, respectively}.
    \label{fig:setup}
\end{figure}

The phase-field simulation is set up and performed in a 2-D half-cell configuration, as illustrated in Figure \ref{fig:setup}. The initial polycrystalline microstructure of NAS in the simulation is generated using a phase-field model of grain growth\cite{krill2002computer}, which gives $\phi_{g}$ distributions. The left boundary of the simulation region is the Na anode. We apply a time-dependent electric potential on the left and right boundaries:

\begin{equation}
    \phi(t) = 
    \begin{cases}
        \quad -\phi_0 \cdot \tanh\left[10\sin\left(\frac{2\pi t_{\rm total}}{t_{\rm cycle}}\right)t\right], x=0\\
        \quad 0, x=1
    \end{cases}
\end{equation}

where $\phi_0$ is set to 0.2~V in this study. $t_{\rm total}$=0.01 s and $t_{\rm cycle}$=0.002 s are the total simulation time and the time for each plating-stripping cycle, respectively. $x=0$ and $x=1$ are the left and right boundaries of the system, representing the anode and solid electrolyte, respectively. Dirichlet boundary conditions are imposed on $\xi$ and $\tilde{c}_{\mathrm{Na}^+}$. Specifically, at $x=0$, $\xi=1$ and $\tilde{c}_{\mathrm{Na}^+}=0$, while at $x=1$, $\xi=0$ and $\tilde{c}_{\mathrm{Na}^+}=1$ as Figure \ref{fig:setup} indicates.

 All simulations were performed using the finite element method using COMSOL Multiphysics 6.2. The simulation domain size is  $200 \times 200$ $\mu$m$^2$, discretized into a $250 \times 200$ rectangular mesh. The key parameters used in the phase-field simulations are summarized in Table \ref{tab:phasefield_params}.


\begin{table}[!h]
  \centering
  \caption{Key phase-field parameters and their values.}
  \label{tab:phasefield_params}
  \small
  \begin{tabularx}{\textwidth}{@{} X | c X | X X | X @{}}
    \toprule
    \textbf{Parameter} & \textbf{Symbol} & \textbf{Real value} & \textbf{Normalization} &\textbf{Reduced value}  &\textbf{Notes} \\
    \midrule
    Interfacial mobility (Na)      & $L_{\xi}$     & \SI{2.5e-6}{m^3/(J\times s)} &$\tilde{L_{\xi}} = L_{\xi}\times (E_0 \times \Delta t_0)$ &\SI{1.5e5}{} &\\
    Reaction prefactor             & $L_{\phi}$    &   \SI{1.0}{/s}    &  $\tilde{L_{\phi}} = L_{\phi}\times \Delta t_0$   &4000 &    \\
    Gradient energy coefficient    & $\kappa$&  \SI{3.68e-6}{J/m}& $\tilde{\kappa} = \kappa/(E_0\times l_0^2)$ & \SI{2.45e-5}{} &\\
    Interface width                & $\lambda$ &   \SI{2.5e-6}{m} & $\tilde{\lambda} = \lambda/l_0$ &  \SI{2.5e-2}{}&   \\
    Interfacial energy & $\gamma$  &    \SI{0.98}{J/m^2} &  $\tilde{\gamma}= \gamma/(E_0\times l_0)$&  \SI{6.5e-4}{} &   (approx.) \\
    Barrier height                 & $w$   &   \SI{3.36e6}{J/m^3}  &  $\tilde{w}= w/E_0$  & 0.16 &   \\
    Time step                      & $\Delta t$ &   \SI{0.2}{s} &  $\Delta \tilde{t} = \Delta t/\Delta t_0$   &\SI{5e-5}{} &\\
    Domain size                    & $l$ &  \SI{2.0e-4}{m} &  $\tilde{l} = l/l_0$  &2 &\\
    Electrical conductivity (Na)          & $\sigma^e_{\rm Na}$ & \SI{2.1e7}{S/m} &  $\tilde{\sigma}^e_{\rm Na} = \sigma^e_{\rm Na}\times\frac{\Delta t_0RT}{l_0^2c_0F^2}$ &\SI{1.21e9}{} & \\
    Ionic conductivity (SE)   & $\sigma^i_{\rm SE}$  &  \SI{3.1e-5}{S/m} & $\tilde{\sigma}^i_{\rm SE} = \sigma^i_{\rm SE}\times\frac{\Delta t_0RT}{l_0^2c_0F^2}$ &\SI{1.78e-3}{} &\cite{gamo2019multiphase} \\
    Diffusivity (in Na)                & $D_{\rm Na}$ &   \SI{1.15e-16}{m^2/s}&   $\tilde{D}_{\rm Na} = D_{\rm Na}\times(\Delta t_0/l_0^2)$    &\SI{4.6e-5}{}        & assume same as $D_{\rm SE}$    \\
    Diffusivity (in \ce{Na3Sb})        & $D_{\ce{Na3Sb}}$   &  \SI{1.15e-20}{m^2/s}&  $\tilde{D}_{\ce{Na3Sb}} = D_{\ce{Na3Sb}}\times(\Delta t_0/l_0^2)$   &\SI{4.6e-9}{}        & assume $10^{-4}\times D_{\rm Na}$\\
    Diffusivity (in SE)                & $D_{\rm SE}$ &     \SI{1.15e-16}{m^2/s}&   $\tilde{D}_{\rm SE} = D_{\rm SE}\times(\Delta t_0/l_0^2)$
    & \SI{4.6e-5}{}    &   calculated from\cite{gamo2019multiphase}\\
    \midrule
    \multicolumn{6}{@{} p{\textwidth} @{}}{\footnotesize
    \textbf{Note:}
    $l_0 = $\SI{1e-4}{m}, 
    $E_0=$\SI{1.5e7}{J/m^3},
    $\Delta t_0=$ \SI{4e3}{s},
    $c_0=$\SI{7.22e4}{mol/m^3}
}
\\    
  \end{tabularx}
\end{table}

\section{3. Results and Discussions}

\subsection{3.1 Surface Electron Density}

In this section, we present the DFT-calculated surface electron densities and a corresponding parametric study to evaluate the sensitivity of our model to these values and their inherent anisotropy. Our findings indicate that while the absolute magnitude of the electron density is a critical parameter, the calculated anisotropy has a negligible impact on the overall results. This justifies the use of an isotropic approximation in subsequent simulations, simplifying the model without sacrificing physical accuracy.

\subsubsection{3.1.1 DFT Results for Surface Electron Density}

\begin{figure}[!h]
    \centering
    \includegraphics[width=\linewidth]{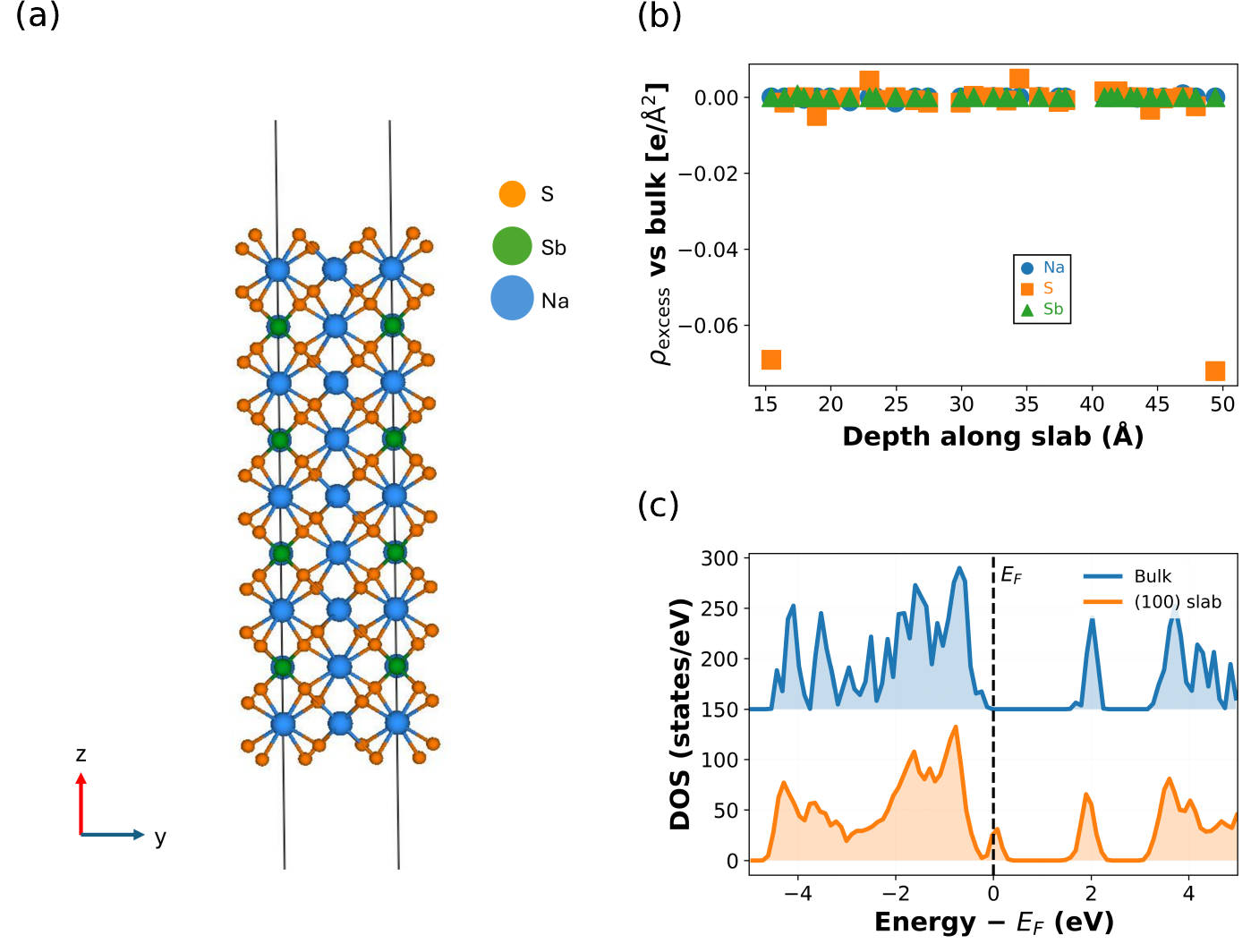}
    \caption{(a) NAS slab of (100) direction with elements marked \ce{Na} (blue), \ce{S} (orange), and \ce{Sb} (green). (b) DFT results for excess electron charge densities from Bader charges for the NAS (100) surface. (c) Electronic density of states (DOS) of the NAS (100) slab in comparison to the bulk DOS.}
    \label{fig:100set}
\end{figure}

Periodic symmetric slabs of NAS  for the (100), (110), and (111) surfaces with at least {35~\AA} vacuum layer were constructed and relaxed with dipole corrections enabled to correct the errors introduced by interslab interactions through the periodic boundary conditions. Figure \ref{fig:100set}(a) shows the (100) surface as an example. Surface charges on the atoms were calculated from Bader charge analysis\cite{henkelman2006fast, sanville2007improved,tang2009grid} where the Bader charges of the different species in the relaxed bulk structure were subtracted from their equivalents in the slab. Figure \ref{fig:100set}(b) shows the resulting excess charges for the (100) slab with $T=300$~K Fermi smearing. The outermost surface atoms are located at 16 \text{\AA} and 50 \text{\AA}, respectively. The corresponding excess charges for the (110) and (111) surfaces are shown in Figures S1 and S2 in the Supplementary Information. For the (100) and (110) slabs, surface dipoles arise primarily from dangling \ce{S} bonds, which are the primary sites for charge redistribution. As is well known,\cite{Windl04} these unsaturated bonds lead to deep gap-states, which can be seen in Figure \ref{fig:100set}(c) in the electronic Density of States (DOS) of the slab in comparison to the bulk DOS. The fact that the Fermi level for the slab DOS cuts through the defect peak indicates that the gap state from the dangling bond is occupied by an electron not involved in covalent bonding, which thus results in surface charge.

Based on the DFT calculations, we determined the surface electron densities for different NAS crystal planes, as summarized in Table \ref{tab:DFT_electron}. The surface electron density exhibits some anisotropy, where the value on the (100) plane is approximately 2.6 and 2.3 times higher than those on the (110) and (111) planes, respectively. This variation can be attributed to differences in the surface S and Sb-atom density and the resulting dangling-bond charges as shown in Figs. S1 and S2 in the SI. 

While this anisotropy can be mathematically described by Eq.\ S3 and integrated into the phase-field model, our sensitivity analysis in the following section and in Section S3 of the Supplementary Information indicates that using anisotropic charge density yields only small differences compared to an isotropic approximation. Consequently, we assume an orientation-independent surface electron concentration, calculated by averaging over the sampled crystallographic orientations. This simplification is particularly advantageous given the structural complexity of GBs in SEs, where identifying  individual planes and evaluating high-angle GBs remains a significant challenge.  

The isotropic surface electron density used in our phase-field simulations is $c_{\rm surf} = {\rho_{e^-}^{\rm avg}}/({2 N_A d}) = 5.89\times10^{-2}$~mol/cm$^3$, where $\rho_{e^-}^{\rm avg} = 7.09\times10^{14}$~${\rm cm}^{-2}$ is the Boltzmann-weighted average for the surface charge including the facet multiplicities as described in Sec.~ S2 in the SI, shown in Table \ref{tab:DFT_electron}. Furthermore, $N_A$ is the Avogadro constant, and $d=1$~{\AA} is the approximate surface depth of excess electrons observed in our results such as shown in Fig.\ \ref{fig:100set}(b). Alternative averaging approaches commonly found in the literature where also examined as discussed in Sec. S2 of the SI. Since all the average values vary by less than 10\%, the final surface electron concentration is taken from the Boltzmann-weight average with facet multiplicities. 

Details of the surface energy calculations\cite{sen2007,mishra2012native,wang2021native, rush2017unraveling} and the associated chemical potentials applied in the Boltzmann average calculations are provided in Sec.~S2 of the Supplementary Information: "Additional average surface electron concentration and surface energy calculations". Table \ref{tab:DFT_electron} summarizes the surface electron concentration and the surface energy ($\gamma$) from all three (100), (110), and (111) planes. Details of the anisotropic surface electron concentration application are also provided in the Sec. S3 of the SI: "Anisotropic surface electron concentration".

\begin{table}[!h]
  \centering
  \caption{Excess surface electron densities and surface energies of NAS for (100), (110), and (111) planes. }
  \label{tab:DFT_electron}
  \small
  \begin{tabularx}{\textwidth}{@{} l c c c c c  @{}}
    \toprule
    \rule[-1.8ex]{0pt}{5.4ex}
    \textbf{Plane} &
    \makecell{\textbf{z length} \\ \textbf{(\r{A})}} &
    \makecell{\textbf{$\gamma$} \\ \textbf{(meV/\AA$^2$)}}   &
    \makecell{\textbf{Surface e$^-$ density} \\ \textbf{({cm}$^{-2}$)}} &
    \makecell{\textbf{ Average~$\rho_{e^-}^{\rm avg}$} \\ \textbf{({cm}$^{-2}$)}} &
    \makecell{\textbf{Location of}\\ \textbf{excess electrons}} \\
    \midrule
    100      & 71.67  & 0.041 &$1.47\times10^{15}$ &   & S\\ 
    110      & 81.06  & 0.021 &$5.69\times10^{14}$ &  $7.09\times10^{14}$  & S\\ 
    111      & 74.48  & 0.027 &$6.41\times10^{14}$ &    & Sb\\ 
    \bottomrule
  \end{tabularx}
\end{table}

\subsubsection{3.1.2 Parametric Phase-Field Study for Surface Electron Density}

In this section, we provide justifications for incorporating the isotropic surface electron concentration obtained from DFT calculations into the phase-field framework. Firstly, we justify approximating GB electron concentrations using surface electron concentrations. In realistic cold-pressed or sintered \ce{Na3SbS4} electrolytes, pores and grain boundaries coexist with broadly distributed surface orientations and GB misorientations. Quantifying the electron concentration of general high-angle GBs directly by DFT remains challenging because constructing representative GB supercells requires substantially larger cells and computational cost, while more complex atomistic approaches may be needed for realistic disordered interfaces. Similar surface-to-GB electron-density approximations have also been adopted in previous studies.\cite{tian2019interfacial}

Secondly, in order to demonstrate {\it a posteriori} why the DFT calculation of the charge density was necessary, we performed a parameter study of the surface electron concentration for one plating-stripping cycle, as shown in Supplementary Figure S5. The simulation results clearly demonstrate that the surface electron concentration critically determines both the Na dendrite penetration depth during plating, and the size of isolated Na (often referred to as “dead” Na) during stripping, as detailed in Sec.~S4 of the Supplementary Information. 

Thirdly, to justify the incorporation of isotropic averaged surface electron concentrations, we also performed a phase-field simulation with anisotropic surface electron concentrations in the 2-D xy-plane, with details reported in Sec.~S3 of the Supplementary Information. The results demonstrate that the overall dendrite penetration depth remains nearly identical to that obtained using the isotropic surface electron concentration.The primary difference is that the dendrite tip morphology exhibits a slight orientation preference compared with the isotropic case. These results support the use of the averaged isotropic surface electron concentration described in the previous section for all the following simulations. Future work will explicitly quantify GB-specific electron densities using DFT and complementary atomistic methods. 

In addition, although other Na solid electrolytes were not explicitly simulated in this work, previous DFT calculations have shown that surfaces and grain boundaries can reduce the electronic bandgap in several Na solid electrolytes, including \ce{Na2S}, \ce{Na3PS4}, and NASICON.\cite{xie2024effects} These results suggest that enhanced surface- or GB-associated electronic states may be a broader feature across Na solid electrolytes, although the magnitude of the effect depends strongly on both the material chemistry and the specific surface or GB orientation. Therefore, extension of the present DFT–phase-field framework to other sulfide solid electrolytes is indeed feasible to further investigate the Na dendrite growth and dead Na accumulation behaviors in different SEs.

\subsection{3.2 Isolated Metallic Na Formation and its Impact on Continuous Cycling}

\begin{figure}[!h]
    \centering
    \includegraphics[width=\linewidth]{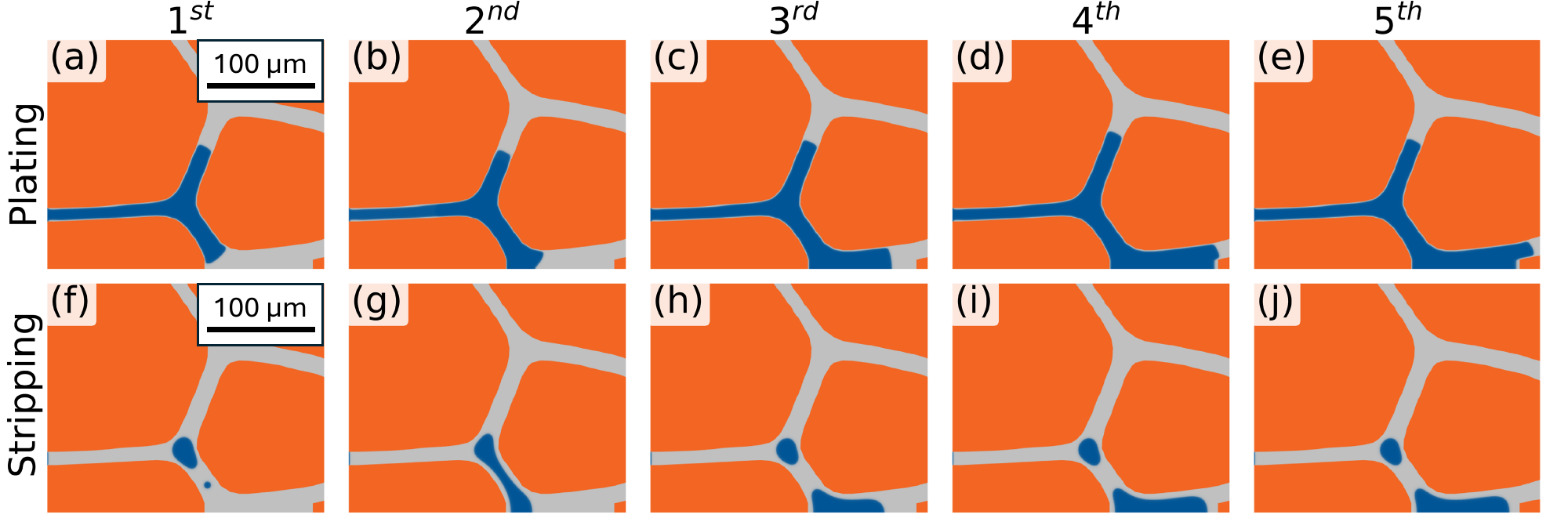}
    \caption{Snapshots of the Na dendrite(blue)– polycrystalline NAS SE(orange)- GBs(gray) during five consecutive electrochemical cycles under an applied voltage of 0.2 V. Panels (a–e) show the plating process, and panels (f–j) show the corresponding stripping process. The sequence highlights the role of isolated Na remnants formed during stripping, which act as preferential sites for subsequent Na deposition and progressively promote dendrite growth over repeated cycling. Full cycling animation is provided in Supplementary Figure S8.}
    \label{fig:Na_0.2V_set}
\end{figure}

In this section, we investigate the Na plating and stripping behaviors inside grain boundary channels connected to the Na anode, with 5 consecutive plating-stripping cycles at an applied voltage of $\pm$0.2~V (Figure \ref{fig:Na_0.2V_set}).

Figure \ref{fig:Na_0.2V_set}(a) and (f) show that dead Na already forms at the grain triple junction of the solid electrolyte during the first stripping cycle. During subsequent plating, the advancing dendrite from the anode preferentially reconnects with these isolated Na regions, reactivating them and re-establishing electronic transport. Once reconnected, electrons can again be supplied to the dendrite front, enabling continued Na penetration into the solid electrolyte. Consequently, isolated Na not only facilitates further dendrite growth during plating but also exhibits distinct kinetic behaviors during stripping.

\begin{figure}
    \centering
    \includegraphics[width=0.5\linewidth]{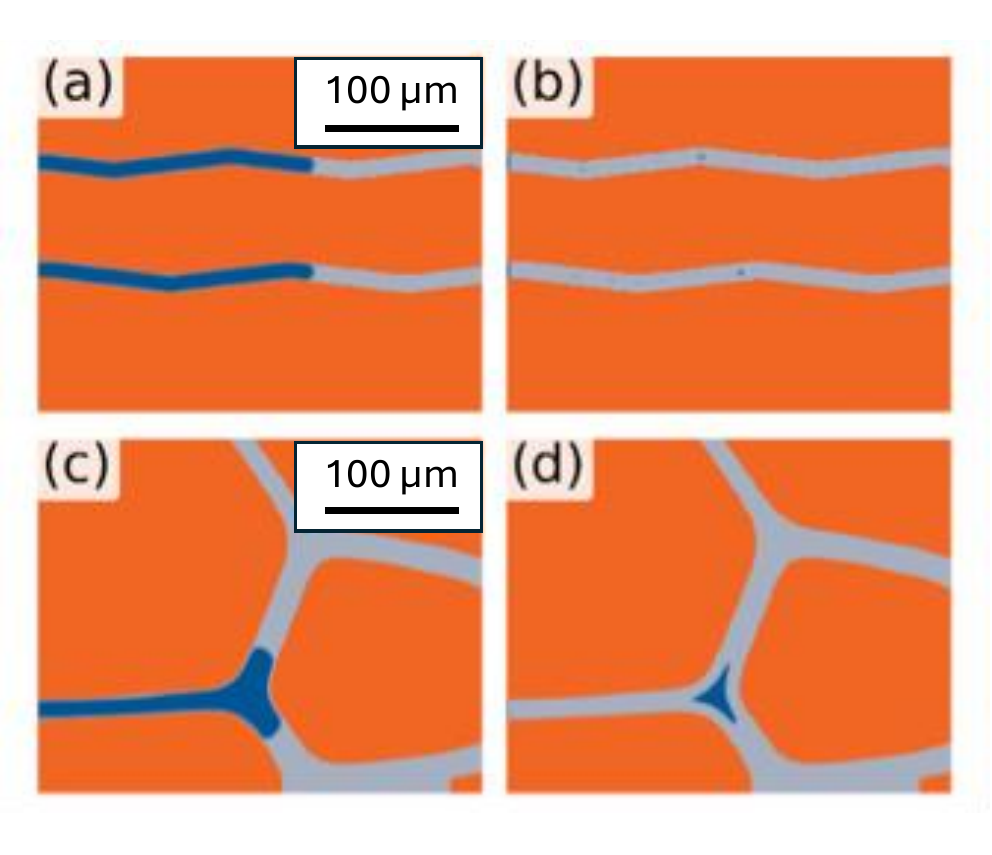}
    \caption{Sensitivity analysis of the governing factors for isolated Na formation within SE grain boundaries. (a-b) Na plating and stripping in parallel GBs without triple junctions, isolating the impact of GB geometry and confinement on metal propagation. (c-d) Na plating and stripping in a hypothetical system with uniform conductivity, demonstrating that without local electronic/ionic heterogeneities, isolated metallic Na formation is significantly suppressed.}
    \label{fig:sensitivity}
\end{figure}

\begin{figure}[!h]
    \centering
    \includegraphics[width=\linewidth]{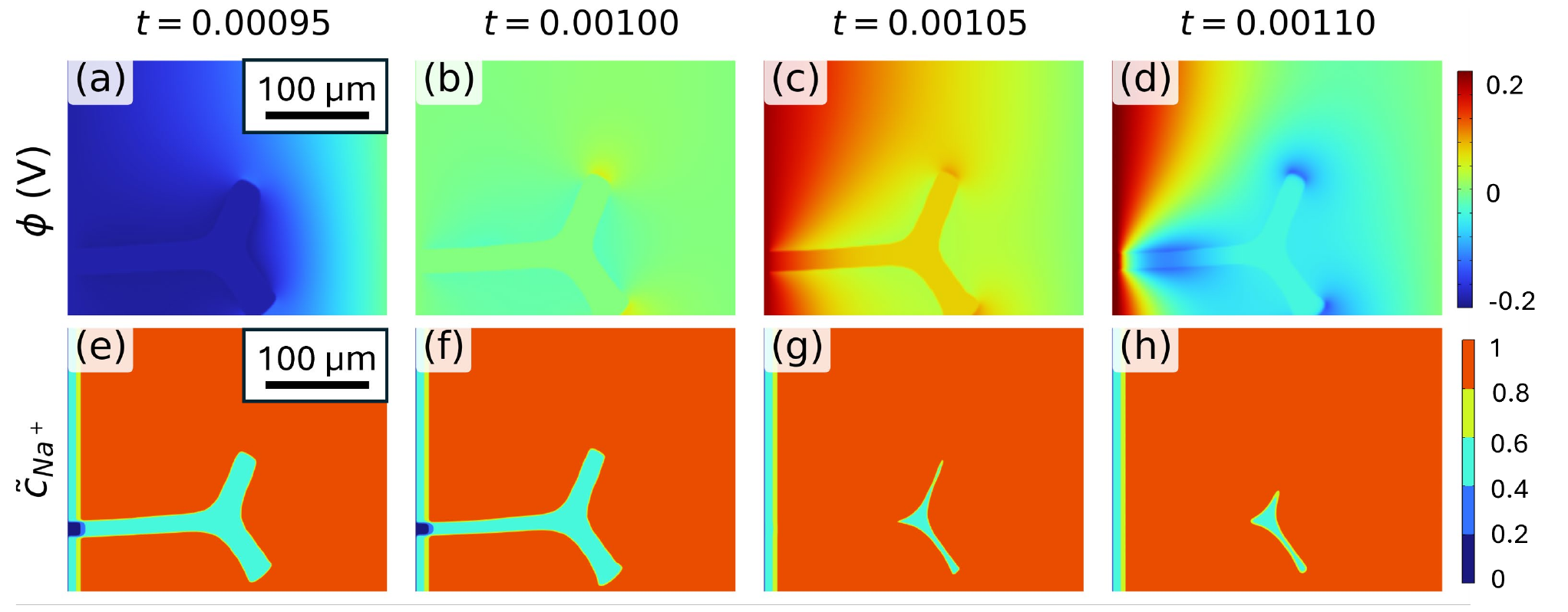}
    \caption{Electric potential (a–d) and \ce{Na^+} concentration (e–h) distributions immediately before and after dendrite stripping. Early stripping initiates along the Na–grain-boundary sidewalls due to sharp potential gradients (a,b). Following electronic disconnection, a localized negative-potential basin forms around the isolated Na, indicating field redistribution and current bypass (c,d). The corresponding \ce{Na^+} shielding and curvature contrast between the narrowed neck and the triple-junction Na highlight transport-limited dissolution and kinetic stabilization of isolated Na (e–h). }
    \label{fig:ele and ion}
\end{figure}

The root cause of isolated Na formation can be attributed to asymmetric electronic accessibility imposed by the SE GB geometry. At the onset of stripping, Na dissolution proceeds primarily through rapid sidewall retreat. This behavior arises from the sharp potential gradient between the Na dendrite inside GB and the surrounding SE (Figure \ref{fig:ele and ion}(b)). The early-stage stripping is thus dominated by electrochemical driving force. Consequently, the dendrite progressively thins, as illustrated in Figure \ref{fig:ele and ion}(e-f). As stripping continues, the dendrite stem becomes thinner than the Na region anchored near the grain triple junction. This morphological asymmetry imposes an electronic transport constraint: electrons can only be supplied through a narrowing stem, which acts as a bottleneck. The resulting current constriction localizes dissolution at the stem immediately upstream of the triple junction, producing a necking instability and ultimately reducing the electronic connection. As a result, Na at the triple junction becomes isolated. The sharp curvature developed near the neck in Figure \ref{fig:ele and ion}(g) reflects the strong spatial imbalance in dissolution rate between the constricted stem and the more stable Na region beyond the triple junction. To further verify the geometric asymmetry introduced by the triple junction, Na stripping in a parallel grain structure with the same parameter setup is shown in the Figure \ref{fig:sensitivity}(a-b). The result shows nearly complete dissolution of Na metal, with little to no isolated Na remaining after stripping.

A second notable phenomenon is observed after disconnection, where the dissolution kinetics of connected dendrites and isolated Na differ markedly. Connected dendrites dissolve rapidly because they remain electronically coupled to the bulk Na anode, enabling sustained charge transfer. In contrast, once Na becomes isolated, its dissolution rate decreases substantially. The stripping of isolated Na is governed by three coupled effects: (a) formation of a negative-potential basin, (b) local \ce{Na^+} shielding, and (c) interfacial curvature.

Figures \ref{fig:ele and ion}(c–d) show that, after disconnection, the isolated Na behaves as a floating metallic inclusion. Polarization under the applied field redistributes the local potential, generating a negative-potential basin and reducing the electric-field strength at the Na surface. In addition, because Na metal is highly electronically conductive while NAS is predominantly ionically conductive, the resulting transport mismatch promotes local \ce{Na^+} accumulation near the isolated Na interface. This accumulation further suppresses stripping(Figure \ref{fig:ele and ion}(g–h)).

Although curvature effects might intuitively be expected to accelerate the dissolution of isolated Na, Figures \ref{fig:ele and ion}(g)–(h) indicate that curvature plays only a limited role once stripping becomes transport-limited. At early stages, curvature-driven thermodynamic instability may contribute to Na dissolution while residual electrons are still available within the isolated Na. However, after electronic depletion, a sharp interfacial curvature can persist while electron transport is effectively cut off, leading to a dramatic drop in the dissolution rate. Thus, despite being thermodynamically unstable and possessing a high local chemical potential, isolated Na becomes kinetically trapped and remains largely intact during stripping. To verify the effect of the conductivity difference, we perform simulations under uniform conductivity throughout the system (Figures \ref{fig:sensitivity}(c-d)). Without the conductivity imbalance, the curvature effect becomes the dominant driving force for stripping along all three branches of the junction, resulting in complete dissolution of Na within the SE without the formation of isolated Na.

Ultimately, isolated Na metal directly promotes continuous dendrite penetration into the SE through repeated reactivation during plating and its reduced dissolution kinetics during stripping. As a result, both the dendrite and the isolated Na progressively penetrate deeper into the solid electrolyte over successive cycles, as evidenced by the continuous plating–stripping evolution shown in Figure \ref{fig:Na_0.2V_set}(b)–(e) and (g)–(j).

\subsection{3.3 Na Dendrite Growth through polycrystalline SE particles under varying applied voltage}
\begin{figure}[!h]
    \centering
    \includegraphics[width=\linewidth]{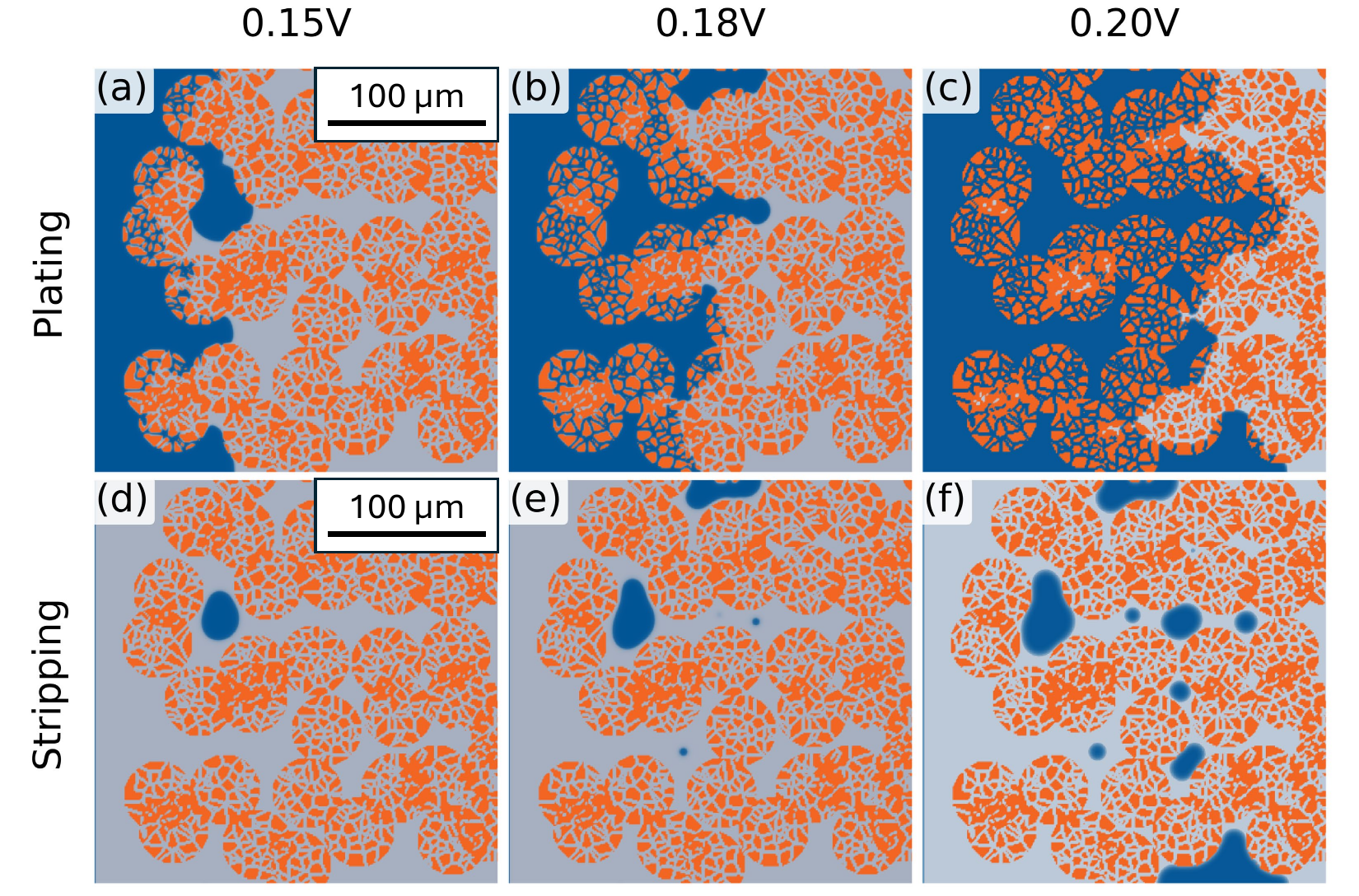}
    \caption{Snapshots of the fifth plating–stripping cycle in a SE microstructure with multiple polycrystalline NAS particles (Na metal (blue), SE (orange), and either void or GBs (gray)) at applied voltages of (a,d) 0.15 V, (b,c) 0.18 V, and (c,f) 0.20 V. Full cycling animation for each applied voltage is listed in Supplementary Figures S9-S11}
    \label{fig:V_set}
\end{figure}

Because isolated Na facilitates continued dendrite penetration into the SE, suppressing its formation may effectively mitigate dendrite growth. To investigate the governing factors of isolated Na formation under more realistic SE microstructures, a NAS SE containing multiple spherical polycrystalline particles with internal GBs is considered (Figure \ref{fig:V_set}). This structure represents a cold-pressed SE microstructure without post-treatment, consistent with typical half-cell fabrication conditions. Applied voltages of 0.15 V, 0.18 V, and 0.20 V are used to evaluate their effects on Na dendrite penetration depth and isolated Na metal formation.

The spherical particle assembly adopted in this study represents an idealized approximation of the microstructure of cold-pressed \ce{Na3SbS4}. Experimental SEM observations show that cold-pressed \ce{Na3SbS4} pellets contain irregularly shaped particles, heterogeneous interparticle contacts, and residual pore space, with substantial changes in packing and densification depending on powder processing and composition.\cite{zhang2021anion} In realistic microstructures, irregular particle shapes and broad grain-size distributions can generate nonuniform grain-boundary spacing and local geometrical constrictions. At the same time, heterogeneous packing can produce preferentially connected grain-boundary and pore networks. In particular, interconnected or elongated pores may provide more continuous free-volume pathways for Na penetration than the isolated pores represented in the present model. They may also increase the likelihood of incomplete Na removal during stripping. Consequently, the exact penetration depth and isolated-Na distribution are expected to depend quantitatively on the actual electrolyte morphology, whereas the present simplified geometry is intended to capture the underlying mechanistic roles of grain boundaries, triple junctions, and pores.

\begin{figure}[!h]
    \centering
    \includegraphics[width=0.5\linewidth]{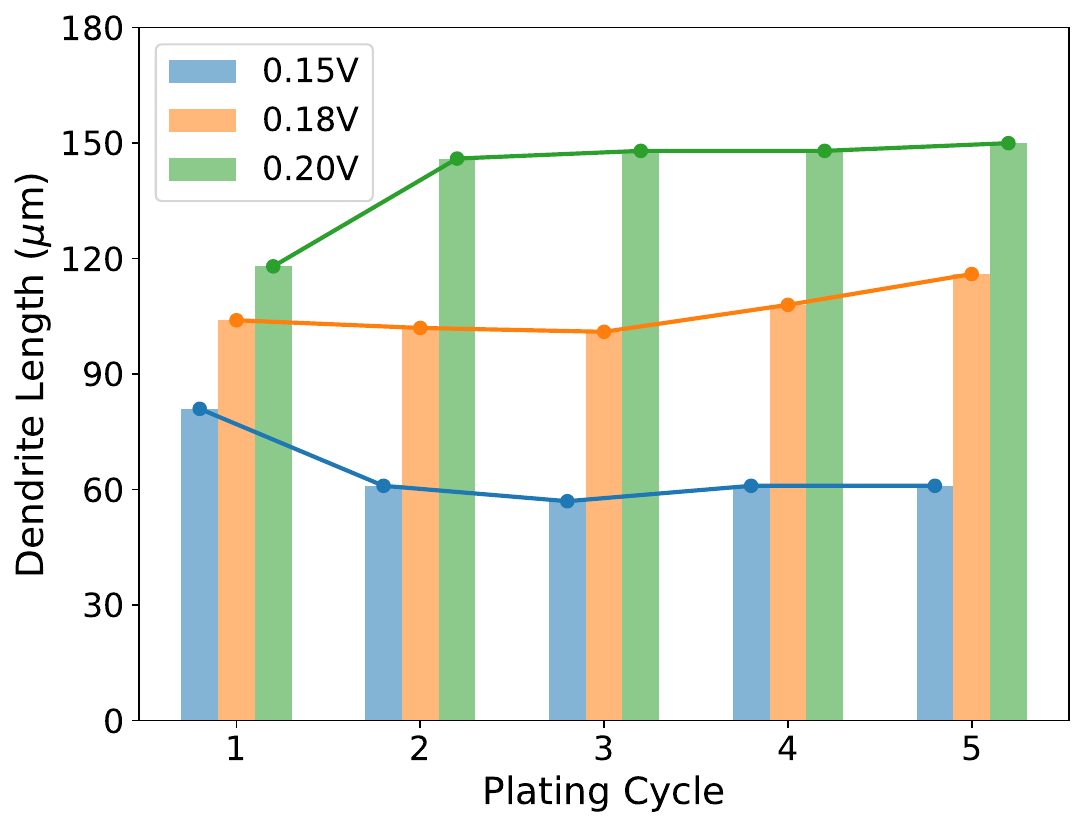}
    \caption{Bar chart summarizing the Na dendrite penetration depth across five consecutive plating–stripping cycles at each applied voltage.}
    \label{fig:barchart_V}
\end{figure}

Figure \ref{fig:V_set} presents snapshots from the fifth plating–stripping cycle, illustrating the voltage-dependent evolution of dendrite morphology and isolated Na. A clear trend is observed in the dendrite penetration depth: the shallowest penetration of 79 $\mu$m occurs at 0.15 V, whereas progressively deeper penetration is observed at 123 $\mu$m and 183 $\mu$m for 0.18 V and 0.20 V, respectively (Figure \ref{fig:barchart_V}). This trend correlates directly with the extent of isolated Na formation, which is minimal at 0.15 V but becomes increasingly pronounced at 0.18 V and 0.20 V within intergranular voids.

The presence of intergranular voids inside the SE microstructure also leads to distinct plating behaviors. Figure \ref{fig:V_set} shows that Na propagates more rapidly within voids than along intragranular GBs. Compared with GBs, voids possess larger surface areas with more surface electrons, enabling rapid Na propagation once an electronically conductive pathway from the anode is established. In addition, voids possess larger free volumes than GBs, thereby imposing fewer geometric constraints on Na dendrite propagation.

In addition to promoting faster dendrite growth, voids also facilitate isolated Na formation more readily than GBs (Figure \ref{fig:V_set}). Rapid Na propagation during plating can also lead to rapid disconnection during stripping, causing Na within voids to lose electronic continuity more easily than Na within GBs. Furthermore, Na deposited in voids is surrounded by SE interfaces, which undergo sidewall stripping as discussed in the previous section. The combined effects of accelerated propagation, enhanced sidewall stripping, and early electronic disconnection lead to the formation of isolated Na within intergranular voids between spherical SE particles. Consequently, at 0.15 V, isolated Na is minimal because Na does not penetrate deeply enough to become enclosed by SE particles. In contrast, at 0.18 V and 0.20 V, sufficient Na deposition occurs within voids, promoting rapid sidewall stripping and the formation of isolated Na. Therefore, as discussed previously, isolated Na acts as a secondary electron source during subsequent plating, thereby assisting further dendrite penetration during cycling.

\subsection{3.4 Impact of SE and Anode Microstructures}
\begin{figure}[!h]
    \centering
    \includegraphics[width=\linewidth]{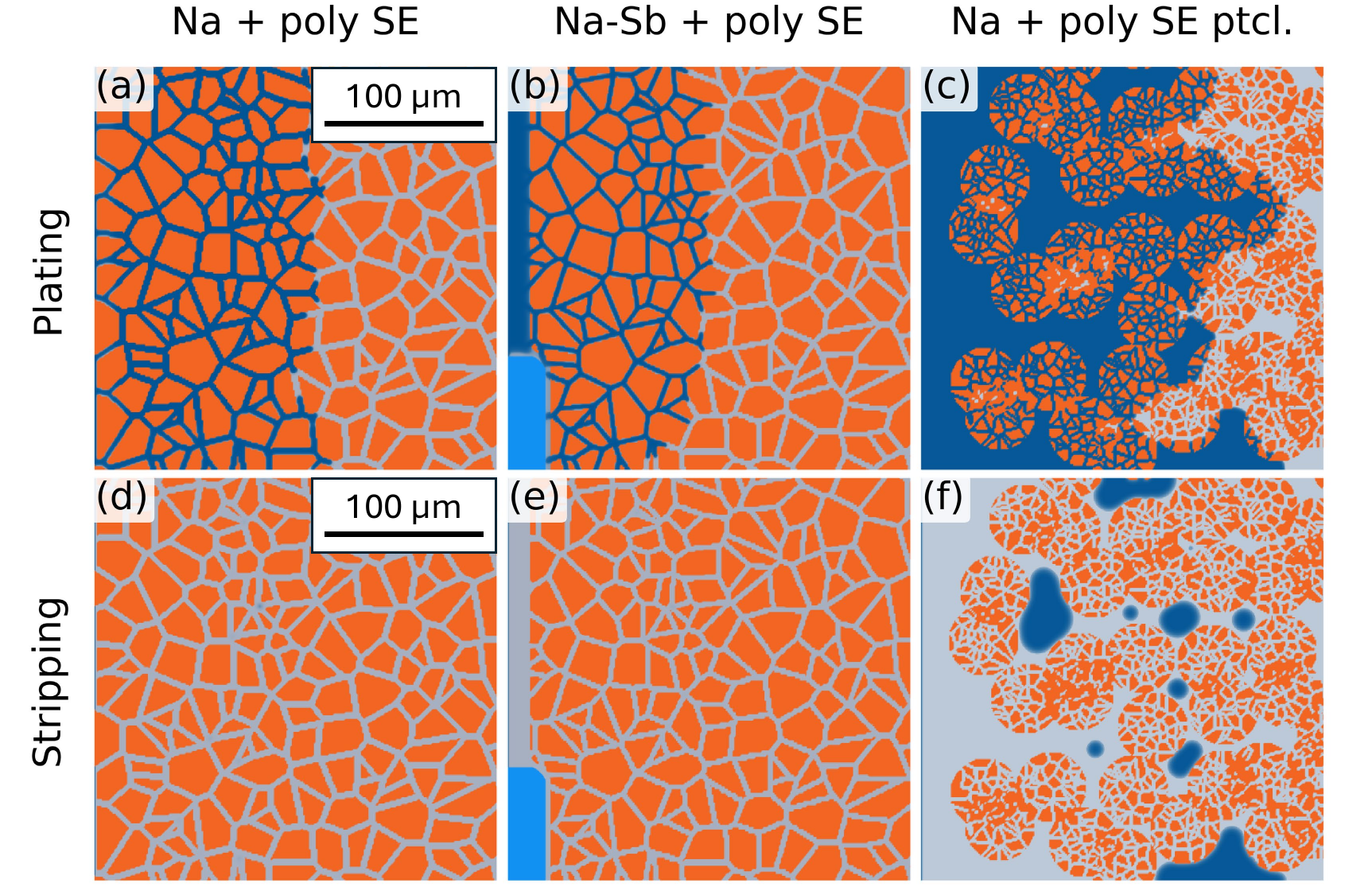}
    \caption{Snapshots of the fifth plating–stripping cycle for different SE and anode microstructures. (a,d) Pure Na anode with refined polycrystalline SE; (b,e) (Na+Na$_3$Sb) composite anode with refined polycrystalline SE; (c,f) Pure Na anode with multiple polycrystalline SE particles. Na metal is shown in blue, SE grains is shown in orange, \ce{Na3Sb} is shown in light blue, and gray indicates either voids or GBs. Additional full cycling animation for pure Na anode and alloyed-anode in dense microstructure is provided in Supplementary Figures S12 and S13.}
    \label{fig:SE_set}
\end{figure}

The influence of SE and anode microstructures on dendrite growth is critical, as geometric features can strongly modulate transport pathways and affect the stripping behavior discussed previously. To further explore these effects, we consider three cases: (1) a refined SE microstructure with an average grain size of 30 $\mu$m and substantially thinner grain boundaries; (2) the refined SE microstructure in (1), combined with a composite anode (Na+Na$_3$Sb); and (3)a SE with polycrystalline particles and voids as in the previous section. We consider the composite anode (Na+Na$_3$Sb) is that anode engineering has been experimentally explored as a complementary strategy to suppress dendrite penetration. In our simulation, we introduce a single \ce{Na3Sb} particle into the Na anode, which is described by a non-evolving order parameter $\phi_{\ce{Na3Sb}}$. We assume that the \ce{Na3Sb} particle is electronically conductive but has a low ionic diffusivity of $D_{\ce{Na3Sb}} = 10^{-4}D_{\rm Na}$. An applied voltage of 0.20 V was used for all cases (Figure \ref{fig:SE_set}).

\begin{figure}[!h]
    \centering
    \includegraphics[width=0.5\linewidth]{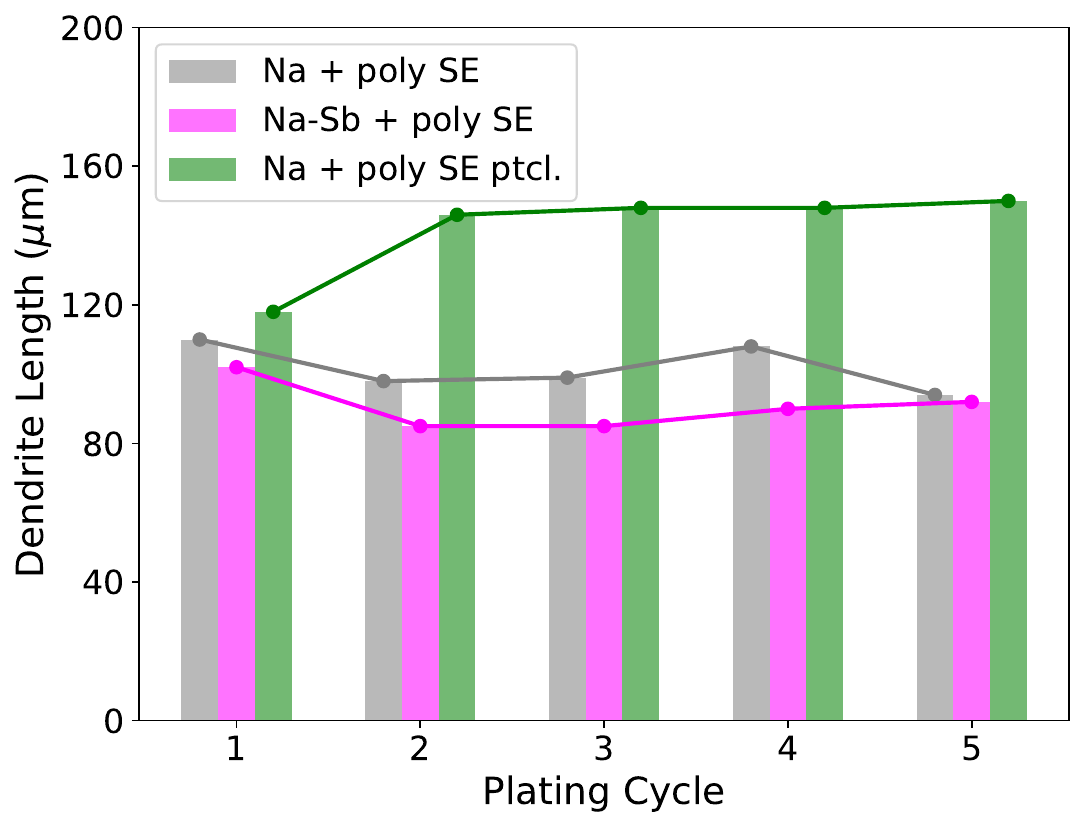}
    \caption{Bar chart summarizing the Na dendrite penetration depth across five consecutive plating–stripping cycles for each microstructural condition.}
    \label{fig:barchartSE}
\end{figure}

Figure \ref{fig:barchartSE} compares the Na metal penetration depth for different SE and anode microstructures. As the grain size decreases, the number of GBs increases, introducing additional transport pathways for both electrons and \ce{Na^+} ions. Moreover, a higher GB density is unfavorable for retaining Na dendrites during stripping because the primary driving force for Na dissolution arises from the potential gradient at the Na metal–SE interface. Consequently, a greater number of GBs increases the interfacial contact area between Na metal and the SE, thereby making interfacial potential gradients more dominant than curvature effects during stripping.

\begin{figure}[!h]
    \centering
    \includegraphics[width=\linewidth]{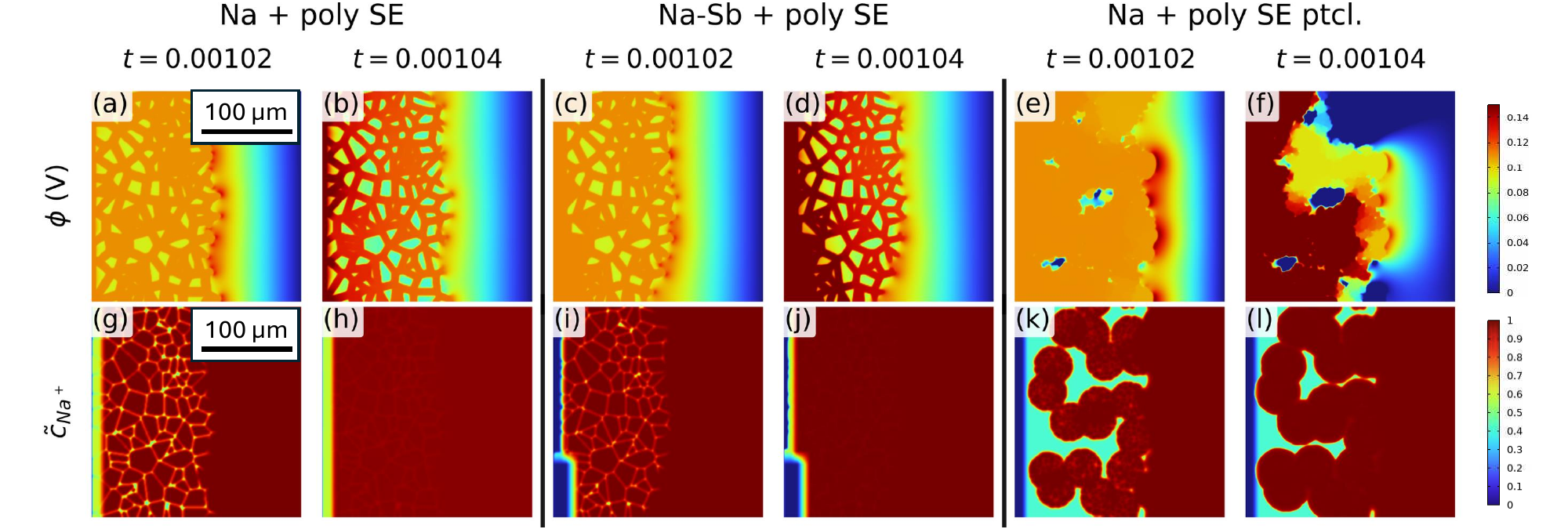}
    \caption{Electric potential and \ce{Na^+} distribution of dense SE (a-b,g-h), dense SE with \ce{Na3Sb} alloyed-anode (c-d,i-j), and polycrystalline SE particles (e-f,k-l) right after the start of stripping, respectively. A different stripping rate is indicated through (h,j,l), showcasing the impact of SE microstructure with GBs and voids.}
    \label{fig:e+i_SE}
\end{figure}

The potential distributions at two early time points during the first stripping cycle, shown in Figure \ref{fig:e+i_SE}(a–f), further illustrate the differences among SE microstructures. For the dense SE, the behavior is similar to that discussed previously (Figure \ref{fig:ele and ion}); however, the presence of more numerous, thinner GBs ensures more complete Na dissolution by providing stronger potential gradients and more effective transport pathways. As a result, isolated Na does not form at triple junctions despite the increased geometric complexity.

From the perspective of anode engineering, introducing an alloyed anode with significantly lower \ce{Na^+} diffusivity suppresses Na deposition during plating, leading to a lower Na concentration in the deposited metal and reduced dendrite penetration from 113 $\mu$m(pure Na anode) to 104 $\mu$m, corresponding to an $8\%$ reduction, after five cycles (Figure \ref{fig:barchartSE}). This effect also lowers the potential gradient between the Na metal and SE at the dendrite front (Figure \ref{fig:e+i_SE}(a,c)), while maintaining sidewall stripping. This behavior is further supported by the \ce{Na^+} concentration snapshots in Figure \ref{fig:e+i_SE}(i,j), compared with the dense SE case using a pure Na anode (Figure \ref{fig:e+i_SE}(g,h)). Together, these effects result in faster Na dissolution within the same stripping duration, highlighting the advantage of the alloyed anode for dendrite suppression and isolated Na mitigation.

In contrast, the previously discussed polycrystalline SE particle microstructure exhibits the deepest dendrite penetration of 183 $\mu$m (Figure \ref{fig:barchartSE}) and a slower dissolution rate during stripping (Figure \ref{fig:e+i_SE}(k,l)) compared with dense SE microstructures. Owing to the presence of voids and a lower GB density, the overall potential gradient in the spherical SE is more spatially homogeneous (Figure \ref{fig:e+i_SE}(e,f)). Consequently, sidewall stripping proceeds more slowly than in dense SEs, where additional GBs provide increased interfacial contact. This contrast is evident when comparing Figures \ref{fig:e+i_SE}(g,h) and \ref{fig:e+i_SE}(k,l): at the same stripping time, a substantial amount of Na metal remains within the spherical SE, consistent with the formation of isolated Na discussed earlier.

Comparing to experimental observation, direct operando visualization of Na penetration at individual grain boundaries and triple junctions in \ce{Na3SbS4} remains experimentally challenging. Recent operando SIMS characterization of solid-state Na cells has revealed Na transport columns extending through ceramic solid electrolytes and linked their formation to bulk/grain-boundary transport, interfacial degradation, and dendrite formation, although the individual grain boundaries could not be directly resolved at the available spatial resolution.\cite{sukumaran2025probing} Thus, the comparison between experiment and the present simulations is primarily mechanistic rather than a direct morphology-to-morphology validation. The experimentally observed preferential Na transport and dendrite formation through heterogeneous electrolyte regions are consistent with the defect-assisted propagation predicted here. In this context, the phase-field simulations provide complementary insight by resolving the temporal evolution of Na along individual grain boundaries, pores, and triple junctions, which remains difficult to capture experimentally.

In summary, the SE microstructure plays a critical role in regulating Na dendrite growth. Increasing the density of fine-GBs provides additional Na metal–SE interfacial area, which enhances sidewall dissolution and reduces the time required for complete stripping. Incorporating a diffusivity-hindered \ce{Na3Sb} particle into the Na anode further suppresses dendrite penetration by lowering the effective \ce{Na^+} transport rate, resulting in reduced Na accumulation within the deposited metal and more efficient stripping. In contrast, the presence of voids promotes preferential Na deposition while providing a relatively limited Na metal–SE interfacial area, leading to deeper dendrite penetration and the formation of isolated Na metal.

\section{4. Conclusion}
We have reported an atomically-informed phase-field investigation of Na dendrite evolution during multiple plating–stripping cycles in Na SSBs. 

We have first shown with a parametric phase-field study that the interfacial excess electron concentration is a critical parameter that determines the Na dendrite penetration depth during plating and the size of isolated Na  during stripping. Since the electron density is difficult to measure experimentally, we have used DFT calculations to quantify the interfacial electron density, approximated by the surface charges of NAS for three high-symmetry surfaces. A second parametric phase-field study then showed that the observed anisotropy of the surface charge has a negligible effect on the phase-field results, allowing the surface charge density to be approximated isotropically using an averaged value of $7.09\times 10^{14}$~cm$^{-2}$. Incorporation of this parameter into phase-field simulations then enables a mechanistically consistent description of dendrite evolution across cycles.

The phase-field simulations reveal that dendrite stripping is intrinsically asymmetric, leading to the formation of isolated Na metalthat persist between cycles and act as a structural memory for dendrite penetration in subsequent cycles. This isolation originates from geometric constraints at grain-boundary junctions, which promote premature electronic disconnection during stripping. Although thermodynamically unstable, isolated Na metal remains kinetically trapped and is readily reactivated during subsequent plating, accelerating dendrite penetration with continued cycling.

We further identify applied voltage, electrolyte microstructure, and anode chemistry as effective external controls for suppressing this memory effect. Lower voltages, dense microstructures with fine grain boundaries, and the introduction of a low-diffusivity \ce{Na3Sb} phase collectively enhance stripping efficiency and suppress dendrite growth. Together, these results establish isolated Na as a central origin of dendrite memory effects and provide design principles for stabilizing Na SSBs, with predictions that can be examined experimentally. 

Future experimental validation may employ synchrotron X-ray tomography or in situ electron microscopy to track Na accumulation, void evolution, and residual Na near grain boundaries and triple junctions. The present DFT–phase-field framework can also be extended to other sulfide solid electrolytes, such as \ce{Na3PS4}-based and compositionally modified \ce{Na3SbS4} systems, by incorporating material-specific electronic, transport, and interfacial parameters obtained from first-principles calculations and experiments. Both side reactions and mechanical effects may also be incorporated in future work to capture the coupling between chemo-electrochemical evolution and stress development in SSBs. Such extensions would enable systematic comparison of how electrolyte chemistry, microstructure, and grain-boundary properties influence Na penetration and isolated-Na formation.

\section{Acknowledgement}
The authors acknowledge the start-up fund from The Ohio State University and the College of Engineering Strategic Initiative Research Grant. This work was also supported in part by The Ohio State University Materials Research Seed Grant Program, funded by the Center for Emergent Materials, an NSF-MRSEC, grant DMR-2011876, the Center for Exploration of Novel Complex Materials, and the Institute for Materials and Materials Research. WW acknowledges partial funding from SPACE-MAT, an AFRL Center of Excellence, AFRL grant number FA8650-24-2-5210. Computations were supported by the Ohio Supercomputer Center. We thank Dr. Jinwoo Hwang for many insightful discussions and suggestions for the work and the manuscript.

\bibliography{main} 

@article{Windl04,
author = {Windl, Wolfgang},
title = {Diffusion in silicon and the predictive power of ab-initio calculations},
journal = {physica status solidi (b)},
volume = {241},
number = {10},
pages = {2313-2318},
doi = {https://doi.org/10.1002/pssb.200404931},
url = {https://onlinelibrary.wiley.com/doi/abs/10.1002/pssb.200404931},
eprint = {https://onlinelibrary.wiley.com/doi/pdf/10.1002/pssb.200404931},
year = {2004}
}

@article{chen2015modulation,
  title={Modulation of dendritic patterns during electrodeposition: A nonlinear phase-field model},
  author={Chen, Lei and Zhang, Hao Wei and Liang, Lin Yun and Liu, Zhe and Qi, Yue and Lu, Peng and Chen, James and Chen, Long-Qing},
  journal={Journal of Power Sources},
  volume={300},
  pages={376--385},
  year={2015},
  publisher={Elsevier}
}

@article{zheng2024thermal,
  title={Thermal state monitoring of lithium-ion batteries: Progress, challenges, and opportunities},
  author={Zheng, Yusheng and Che, Yunhong and Hu, Xiaosong and Sui, Xin and Stroe, Daniel-Ioan and Teodorescu, Remus},
  journal={Progress in Energy and Combustion Science},
  volume={100},
  pages={101120},
  year={2024},
  publisher={Elsevier}
}

@article{janek2023challenges,
  title={Challenges in speeding up solid-state battery development},
  author={Janek, J{\"u}rgen and Zeier, Wolfgang G},
  journal={Nature Energy},
  volume={8},
  number={3},
  pages={230--240},
  year={2023},
  publisher={Nature Publishing Group UK London}
}

@article{vera2023environmental,
  title={Environmental impact of direct lithium extraction from brines},
  author={Vera, Mar{\'\i}a L and Torres, Walter R and Galli, Claudia I and Chagnes, Alexandre and Flexer, Victoria},
  journal={Nature Reviews Earth \& Environment},
  volume={4},
  number={3},
  pages={149--165},
  year={2023},
  publisher={Nature Publishing Group UK London}
}

@article{zhao2018solid,
  title={Solid-state sodium batteries},
  author={Zhao, Chenglong and Liu, Lilu and Qi, Xingguo and Lu, Yaxiang and Wu, Feixiang and Zhao, Junmei and Yu, Yan and Hu, Yong-Sheng and Chen, Liquan},
  journal={Advanced Energy Materials},
  volume={8},
  number={17},
  pages={1703012},
  year={2018},
  publisher={Wiley Online Library}
}

@article{tian2019interfacial,
  title={Interfacial electronic properties dictate Li dendrite growth in solid electrolytes},
  author={Tian, Hong-Kang and Liu, Zhe and Ji, Yanzhou and Chen, Long-Qing and Qi, Yue},
  journal={Chemistry of Materials},
  volume={31},
  number={18},
  pages={7351--7359},
  year={2019},
  publisher={ACS Publications}
}

@article{lee2019sodium,
  title={Sodium metal anodes: emerging solutions to dendrite growth},
  author={Lee, Byeongyong and Paek, Eunsu and Mitlin, David and Lee, Seung Woo},
  journal={Chemical reviews},
  volume={119},
  number={8},
  pages={5416--5460},
  year={2019},
  publisher={ACS Publications}
}

@article{hayashi2019sodium,
  title={A sodium-ion sulfide solid electrolyte with unprecedented conductivity at room temperature},
  author={Hayashi, Akitoshi and Masuzawa, Naoki and Yubuchi, So and Tsuji, Fumika and Hotehama, Chie and Sakuda, Atsushi and Tatsumisago, Masahiro},
  journal={Nature communications},
  volume={10},
  number={1},
  pages={5266},
  year={2019},
  publisher={Nature Publishing Group UK London}
}

@article{wang2012tracking,
  title={Tracking lithium transport and electrochemical reactions in nanoparticles},
  author={Wang, Feng and Yu, Hui-Chia and Chen, Min-Hua and Wu, Lijun and Pereira, Nathalie and Thornton, Katsuyo and Van der Ven, Anton and Zhu, Yimei and Amatucci, Glenn G and Graetz, Jason},
  journal={Nature communications},
  volume={3},
  number={1},
  pages={1201},
  year={2012},
  publisher={Nature Publishing Group UK London}
}

@article{wei2025intermediate,
  title={Intermediate phase induced in situ self-reconstruction of amorphous NASICON for long-life solid-state sodium metal batteries},
  author={Wei, Benben and Huang, Shuo and Wang, Xuan and Liu, Min and Huang, Can and Liu, Ruoqing and Jin, Hongyun},
  journal={Energy \& Environmental Science},
  volume={18},
  number={2},
  pages={831--840},
  year={2025},
  publisher={Royal Society of Chemistry}
}

@article{liu2024situ,
  title={In situ forming NaSn alloy/Na2S interface layer for ultrastable solid state sodium batteries},
  author={Liu, Tinghu and Xiang, Pan and Li, Yunming and Li, Zhendong and Sun, Huazhang and Yang, Jing and Tian, Ziqi and Yao, Xiayin},
  journal={Advanced Functional Materials},
  volume={34},
  number={32},
  pages={2316528},
  year={2024},
  publisher={Wiley Online Library}
}

@article{gamo2019multiphase,
  title={Multiphase Na3SbS4 with high ionic conductivity},
  author={Gamo, Hirotada and Phuc, Nguyen Huu Huy and Matsuda, Reiko and Muto, Hiroyuki and Matsuda, Atsunori},
  journal={Materials Today Energy},
  volume={13},
  pages={45--49},
  year={2019},
  publisher={Elsevier}
}

@article{fuchs2024imaging,
  title={Imaging the microstructure of lithium and sodium metal in anode-free solid-state batteries using electron backscatter diffraction},
  author={Fuchs, Till and Ortmann, Till and Becker, Juri and Haslam, Catherine G and Ziegler, Maya and Singh, Vipin Kumar and Rohnke, Marcus and Mogwitz, Boris and Peppler, Klaus and Nazar, Linda F and others},
  journal={Nature materials},
  volume={23},
  number={12},
  pages={1678--1685},
  year={2024},
  publisher={Nature Publishing Group UK London}
}

@article{zhang2024solid,
  title={Solid-state synthesis improves stability and cycling performance of layered sodium oxide cathodes: A solid-state NMR study},
  author={Zhang, Bao and Ji, Yi and Liang, Lixin and Zheng, Qiong and Chen, Kuizhi and Hou, Guangjin},
  journal={Chemical Engineering Journal},
  volume={485},
  pages={149879},
  year={2024},
  publisher={Elsevier}
}

@article{ren2024visualizing,
  title={Visualizing the SEI formation between lithium metal and solid-state electrolyte},
  author={Ren, Fucheng and Wu, Yuqi and Zuo, Wenhua and Zhao, Wengao and Pan, Siyuan and Lin, Hongxin and Yu, Haichuan and Lin, Jing and Lin, Min and Yao, Xiayin and others},
  journal={Energy \& Environmental Science},
  volume={17},
  number={8},
  pages={2743--2752},
  year={2024},
  publisher={Royal Society of Chemistry}
}

@article{takenaka2014electrolyte,
  title={On electrolyte-dependent formation of solid electrolyte interphase film in lithium-ion batteries: strong sensitivity to small structural difference of electrolyte molecules},
  author={Takenaka, Norio and Suzuki, Yuichi and Sakai, Hirofumi and Nagaoka, Masataka},
  journal={The Journal of Physical Chemistry C},
  volume={118},
  number={20},
  pages={10874--10882},
  year={2014},
  publisher={ACS Publications}
}

@article{ushirogata2015near,
  title={Near-shore aggregation mechanism of electrolyte decomposition products to explain solid electrolyte interphase formation},
  author={Ushirogata, Keisuke and Sodeyama, Keitaro and Futera, Zdenek and Tateyama, Yoshitaka and Okuno, Yukihiro},
  journal={Journal of The Electrochemical Society},
  volume={162},
  number={14},
  pages={A2670},
  year={2015},
  publisher={IOP Publishing}
}

@article{liu2021dendrite,
  title={Dendrite-free lithium based on lessons learned from lithium and magnesium electrodeposition morphology simulations},
  author={Liu, Zhe and Li, Yunsong and Ji, Yanzhou and Zhang, Qinglin and Xiao, Xingcheng and Yao, Yan and Chen, Long-Qing and Qi, Yue},
  journal={Cell Reports Physical Science},
  volume={2},
  number={1},
  year={2021},
  publisher={Elsevier}
}

@article{zhang2025simulating,
  title={Simulating solid electrolyte interphase formation spanning 10 8 time scales with an atomically informed phase-field model},
  author={Zhang, Kena and Ji, Yanzhou and Wu, Qisheng and Nabavizadeh, Seyed Amin and Qi, Yue and Chen, Long-Qing},
  journal={Energy \& Environmental Science},
  year={2025},
  publisher={Royal Society of Chemistry}
}

@article{mu2019numerical,
  title={Numerical simulation of the factors affecting the growth of lithium dendrites},
  author={Mu, Wenyu and Liu, Xunliang and Wen, Zhi and Liu, Lin},
  journal={Journal of Energy Storage},
  volume={26},
  pages={100921},
  year={2019},
  publisher={Elsevier}
}

@article{yurkiv2018phase,
  title={Phase-field modeling of solid electrolyte interface (SEI) influence on Li dendritic behavior},
  author={Yurkiv, Vitaliy and Foroozan, Tara and Ramasubramanian, Ajaykrishna and Shahbazian-Yassar, Reza and Mashayek, Farzad},
  journal={Electrochimica Acta},
  volume={265},
  pages={609--619},
  year={2018},
  publisher={Elsevier}
}

@article{gao2024phase,
  title={Phase field simulation of dendrites morphology evolution in sodium metal batteries},
  author={Gao, Li Ting and Guo, Zhan-Sheng},
  journal={Journal of Power Sources},
  volume={613},
  pages={234961},
  year={2024},
  publisher={Elsevier}
}

@article{gao2024controlling,
  title={Controlling sodium dendrite growth via grain boundaries in Na3Zr2Si2PO12 electrolyte},
  author={Gao, Zhonghui and Bai, Yang and Feng, Junrun and Yang, Jiayi and Liu, Porun and Yuan, Haiyang and Guan, Xuze and Wang, Feng Ryan and Huang, Yunhui},
  journal={Advanced Energy Materials},
  volume={14},
  number={20},
  pages={2304488},
  year={2024},
  publisher={Wiley Online Library}
}

@article{barai2024study,
  title={Study of void formation at the lithium| solid electrolyte interface},
  author={Barai, Pallab and Fuchs, Till and Trevisanello, Enrico and Richter, Felix H and Janek, Jürgen and Srinivasan, Venkat},
  journal={Chemistry of Materials},
  volume={36},
  number={5},
  pages={2245--2258},
  year={2024},
  publisher={ACS Publications}
}

@article{jain2013commentary,
  title={Commentary: The Materials Project: A materials genome approach to accelerating materials innovation},
  author={Jain, Anubhav and Ong, Shyue Ping and Hautier, Geoffroy and Chen, Wei and Richards, William Davidson and Dacek, Stephen and Cholia, Shreyas and Gunter, Dan and Skinner, David and Ceder, Gerbrand and others},
  journal={APL materials},
  volume={1},
  number={1},
  year={2013},
  publisher={AIP Publishing}
}

@article{horton2025accelerated,
  title={Accelerated data-driven materials science with the Materials Project},
  author={Horton, Matthew K and Huck, Patrick and Yang, Ruo Xi and Munro, Jason M and Dwaraknath, Shyam and Ganose, Alex M and Kingsbury, Ryan S and Wen, Mingjian and Shen, Jimmy X and Mathis, Tyler S and others},
  journal={Nature Materials},
  pages={1--11},
  year={2025},
  publisher={Nature Publishing Group UK London}
}

@article{kresse1996efficient,
  title={Efficient iterative schemes for ab initio total-energy calculations using a plane-wave basis set},
  author={Kresse, Georg and Furthm{\"u}ller, J{\"u}rgen},
  journal={Physical review B},
  volume={54},
  number={16},
  pages={11169},
  year={1996},
  publisher={APS}
}

@article{perdew1996generalized,
  title={Generalized gradient approximation made simple},
  author={Perdew, John P and Burke, Kieron and Ernzerhof, Matthias},
  journal={Physical review letters},
  volume={77},
  number={18},
  pages={3865},
  year={1996},
  publisher={APS}
}

@article{kresse1996efficiency,
  title={Efficiency of ab-initio total energy calculations for metals and semiconductors using a plane-wave basis set},
  author={Kresse, Georg and Furthm{\"u}ller, J{\"u}rgen},
  journal={Computational materials science},
  volume={6},
  number={1},
  pages={15--50},
  year={1996},
  publisher={Elsevier}
}

@article{henkelman2006fast,
  title={A fast and robust algorithm for Bader decomposition of charge density},
  author={Henkelman, Graeme and Arnaldsson, Andri and J{\'o}nsson, Hannes},
  journal={Computational Materials Science},
  volume={36},
  number={3},
  pages={354--360},
  year={2006},
  publisher={Elsevier}
}

@article{tang2009grid,
  title={A grid-based Bader analysis algorithm without lattice bias},
  author={Tang, Wei and Sanville, Eric and Henkelman, Gustavo},
  journal={Journal of Physics: Condensed Matter},
  volume={21},
  number={8},
  pages={084204},
  year={2009},
  publisher={IOP Publishing}
}

@article{sanville2007improved,
  title={Improved grid-based algorithm for Bader charge allocation},
  author={Sanville, Edward and Kenny, Steven D and Smith, Roger and Henkelman, Graeme},
  journal={Journal of computational chemistry},
  volume={28},
  number={5},
  pages={899--908},
  year={2007},
  publisher={Wiley Online Library}
}

@article{tian2018computational,
  title={Computational study of lithium nucleation tendency in Li7La3Zr2O12 (LLZO) and rational design of interlayer materials to prevent lithium dendrites},
  author={Tian, Hong-Kang and Xu, Bo and Qi, Yue},
  journal={Journal of Power Sources},
  volume={392},
  pages={79--86},
  year={2018},
  publisher={Elsevier}
}

@article{xu2021guiding,
  title={Guiding the design of heterogeneous electrode microstructures for Li-ion batteries: microscopic imaging, predictive modeling, and machine learning},
  author={Xu, Hongyi and Zhu, Juner and Finegan, Donal P and Zhao, Hongbo and Lu, Xuekun and Li, Wei and Hoffman, Nathaniel and Bertei, Antonio and Shearing, Paul and Bazant, Martin Z},
  journal={Advanced Energy Materials},
  volume={11},
  number={19},
  pages={2003908},
  year={2021},
  publisher={Wiley Online Library}
}

@article{ling2022review,
  title={A review of the recent progress in battery informatics},
  author={Ling, Chen},
  journal={npj Computational Materials},
  volume={8},
  number={1},
  pages={33},
  year={2022},
  publisher={Nature Publishing Group UK London}
}

@article{zhang2019progress,
  title={Progress in 3D electrode microstructure modelling for fuel cells and batteries: transport and electrochemical performance},
  author={Zhang, Duo and Bertei, Antonio and Tariq, Farid and Brandon, Nigel and Cai, Qiong},
  journal={Progress in Energy},
  volume={1},
  number={1},
  pages={012003},
  year={2019},
  publisher={IOP Publishing}
}

@article{liang2012nonlinear,
  title={Nonlinear phase-field model for electrode-electrolyte interface evolution},
  author={Liang, Linyun and Qi, Yue and Xue, Fei and Bhattacharya, Saswata and Harris, Stephen J and Chen, Long-Qing},
  journal={Physical Review E—Statistical, Nonlinear, and Soft Matter Physics},
  volume={86},
  number={5},
  pages={051609},
  year={2012},
  publisher={APS}
}

@article{wang2020application,
  title={Application of phase-field method in rechargeable batteries},
  author={Wang, Qiao and Zhang, Geng and Li, Yajie and Hong, Zijian and Wang, Da and Shi, Siqi},
  journal={npj Computational Materials},
  volume={6},
  number={1},
  pages={176},
  year={2020},
  publisher={Nature Publishing Group UK London}
}

@article{pant2024dendrite,
  title={Dendrite Growth and Dead Lithium Formation in Lithium Metal Batteries and Mitigation Using a Protective Layer: A Phase-Field Study},
  author={Pant, Bharat Raj and Ren, Yao and Cao, Ye},
  journal={ACS Applied Materials \& Interfaces},
  volume={16},
  number={42},
  pages={56947--56956},
  year={2024},
  publisher={ACS Publications}
}

@article{han2025combination,
  title={Combination of high-throughput phase field modeling and machine learning to study the performance evolution during lithium battery cycling},
  author={Han, Dandan and Lin, Chen},
  journal={Energy Storage Materials},
  volume={74},
  pages={103982},
  year={2025},
  publisher={Elsevier}
}

@article{krill2002computer,
  title={Computer simulation of 3-D grain growth using a phase-field model},
  author={Krill III, Carl E and Chen, L-Q},
  journal={Acta materialia},
  volume={50},
  number={12},
  pages={3059--3075},
  year={2002},
  publisher={Elsevier}
}

@article{rush2017unraveling,
  title={Unraveling the electrolyte properties of Na 3 SbS 4 through computation and experiment},
  author={Rush Jr, Larry E and Hood, Zachary D and Holzwarth, NAW},
  journal={Physical Review Materials},
  volume={1},
  number={7},
  pages={075405},
  year={2017},
  publisher={APS}
}

@article{sen2007,
author = {Sen, Dipanjan and Windl, Wolfgang},
year = {2007},
month = {02},
pages = {57-64},
title = {Ab-Initio Study of Energetics of the Si(001)-LaAlO3 Interface},
volume = {4},
journal = {Journal of Computational and Theoretical Nanoscience}
}

@article{mishra2012native,
  title={Native point defects in binary InP semiconductors},
  author={Mishra, Rohan and Restrepo, Oscar D and Kumar, Ashutosh and Windl, Wolfgang},
  journal={Journal of Materials Science},
  volume={47},
  number={21},
  pages={7482--7497},
  year={2012},
  publisher={Springer}
}

@article{wang2021native,
  title={Native point defects from stoichiometry-linked chemical potentials in cubic boron arsenide},
  author={Wang, Yaxian and Windl, Wolfgang},
  journal={Journal of Applied Physics},
  volume={129},
  number={7},
  year={2021},
  publisher={AIP Publishing}
}

@article{zhang2025isolated,
  title={Isolated metallic lithium formation in lithium-metal batteries},
  author={Zhang, Jin and Chadwick, Alexander F and Voorhees, Peter W},
  journal={Cell Reports Physical Science},
  volume={6},
  number={1},
  year={2025},
  publisher={Elsevier}
}

@article{chadwick2025suppression,
  title={Suppression of Dendrites in Metal-Anode Batteries by the Soret and Seebeck Effects},
  author={Chadwick, Alexander F and Zhang, Jin and Swift, Michael W and Chopp, David L and Johannes, Michelle D and Voorhees, Peter W},
  journal={ACS Energy Letters},
  volume={10},
  number={12},
  pages={6281--6287},
  year={2025},
  publisher={ACS Publications}
}

@article{tantratian2021unraveling,
  title={Unraveling the Li penetration mechanism in polycrystalline solid electrolytes},
  author={Tantratian, Karnpiwat and Yan, Hanghang and Ellwood, Kevin and Harrison, Elisa T and Chen, Lei},
  journal={Advanced Energy Materials},
  volume={11},
  number={13},
  pages={2003417},
  year={2021},
  publisher={Wiley Online Library}
}

@article{zhang2021anion,
  title={Anion Control of the Electrolyte Na3--x SbS4--x Br X Extends Cycle Life in Solid-State Sodium Batteries},
  author={Zhang, Xin and Phuah, Kia Chai and Adams, Stefan},
  journal={Chemistry of Materials},
  volume={33},
  number={23},
  pages={9184--9193},
  year={2021},
  publisher={ACS Publications}
}

@article{sukumaran2025probing,
  title={Probing dynamic degradation and mass transport in solid-state sodium-ion batteries using operando simultaneous dual-polarity SIMS},
  author={Sukumaran, Sivakkumaran and Chater, Richard J and Fearn, Sarah and Cooke, Graham and Smith, Noel and Skinner, Stephen J},
  journal={EES Batteries},
  volume={1},
  number={4},
  pages={964--974},
  year={2025},
  publisher={The Royal Society of Chemistry}
}

@article{xie2024effects,
  title={Effects of grain boundaries and surfaces on electronic and mechanical properties of solid electrolytes},
  author={Xie, Weihang and Deng, Zeyu and Liu, Zhengyu and Famprikis, Theodosios and Butler, Keith T and Canepa, Pieremanuele},
  journal={Advanced Energy Materials},
  volume={14},
  number={17},
  pages={2304230},
  year={2024},
  publisher={Wiley Online Library}
}

\end{document}